# Dispersion in Rotating Electron-Positron-Ion Quantum Plasma

Atherv Saxena[1], Sudeep Nadgambe[2], Rajan Mishra[2], Punit Kumar[1*]

[1]*Department of Physics, University of Lucknow, India*

[2]*K. J. Somaiya College of Science and Commerce, University of Mumbai, India*

[1*]*E-mail- punitkumar@hotmail.com*

**Abstract**

Quantum plasmas in astrophysical environments are abundant due to extreme electric, magnetic and gravitational fields. These may be most easily seen in neutron stars, white dwarfs, brown dwarfs, red dwarfs, accretion disk of black holes, pulsars, quasars, etc. In this paper, the propagation of electromagnetic wave in three component e-p-i quantum plasma in a rotating frame has been studied taking into account the particle's spin, Fermi pressure and quantum Bohm potential. Effects specific to this particular environment like rotation as well as gravity have also been included. Dispersion of electron, ion and positron have been obtained separately and their coupling have been done in order to analyse the collective behaviour. It has been observed that quantum effects of Fermi pressure and Bohm potential greatly influence the particle dynamics.

## 1. Introduction

In recent years, there has been a notable interest in the exploration of multi-component quantum plasmas, particularly the electron-positron-ion (e-p-i) plasmas, owing to their occurrence in various astrophysical environments [1,2]. These plasmas, composed of electrons, positrons, and ions, manifest extensively in environments like active galactic cores [3], pulsar magnetospheres [4], black hole accretion discs [5], white dwarf atmospheres [6,7], Van Allen radiation belts [8], also exist in the early universe [9], as well as the centre of our galaxy [10]. The Quantum Hydrodynamic (QHD) model has been harnessed to delineate dissipative e-p-i systems, yielding Kadomtsev-Petviashvili-Burger's equation governing small amplitude ion-acoustic solitary and shock waves [11]. Moreover, the impact of exchange-correlation and pressure degeneracy on the behaviour of Ion Acoustic waves (IAWs) in a heterogeneous dense e-p-i plasma has been investigated [12]. A comprehensive bifurcation analysis of quantum ion-acoustic kink, anti-kink, and periodic waves in densely degenerate plasma encompassing electrons, positrons and positive ions has been carried out [13]. A unique approach by treating electron and positron as distinct fluids, focusing on magnetosonic waves in e-p-i quantum plasma has been developed [14, 15].The distinctive feature of equal mass and charge magnitude differentiates electron-positron (e-p) plasmas apart from conventional electron-ion systems [16, 20]. The influence of ion temperature on the characteristics of large-amplitude ion-acoustic (IA) waves in e-p-i plasma, has been studied [19, 21-23].

In environments with high plasma density, quantum mechanical effects become significant, when the de-Broglie wavelength of charged particles approaches inter-particle spacing. Another crucial condition arises with temperature degeneracy, where the Fermi temperature exceeds the system temperature in such a situation, the plasma behaves like a Fermi gas [35-36]. The distinct attributes of quantum plasma emerge from factors such as the pressure law governing the fermionic nature of charged carriers, quantum effects stemming from electron tunnelling, and the Bohr magnetization linked to the electron's half-spin. Additionally, the quantum Bohm potential introduces alterations in the dispersion of collective modes, particularly at quantum scales. These effects are notably influenced by plasma densities and Fermi temperatures.

In high density quantum plasma, quantum effects and collective behaviour is driven by quantum statistics, and degeneracy pressure [38-41]. The presence of long-range electromagnetic interactions enhances the fluid-like character of the plasma, as the dynamics of charged particles are influenced by collective fields generated by the ensemble [42-44]. In the presence of strong magnetic fields, charged particles exhibit quantized orbital motion (e.g. accretion disks and magnetars), which can be effectively described within hydrodynamic framework [45-47]. Many astrophysical plasmas exhibit a small mean-free path compared to macroscopic scales, which supports the assumption of local equilibrium that is foundation to hydrodynamic models. The QHD framework succeeds in integrating these quantum corrections, especially those arising from the Bohm potential and Fermi pressure, to accurately capture both the collective behaviour and wave dynamics [48-51].The Quantum Hydrodynamic (QHD) model has emerged as a focal point in the investigation of electrostatic and electromagnetic waves within quantum plasmas, because of its straightforward use of macroscopic variables, computational efficiency, and simplicity in handling boundary conditions. [52,53].

This research paper is devoted to the analysis of coupled dispersion relations of multi-component plasmas. Section-2 is devoted to the study of dispersion of electron, positron and ion in a rotating magnetised e-p-i quantum plasma using the QHD model. Section-3 is devoted to the coupling of e-p-i modes. Finally section-4 presents summary and discussion.

## 2. Quantum Plasma Dynamics

We consider a three-component uniform quantum plasma consisting of electrons, positrons and positive ions. The quantum plasma is uniform and immersed in an external magnetic field $B_0(\hat{z})$, and is rotating slowly about an axis making an angle $\theta$ with the $\hat{z}$ axis. The dynamics of a circularly polarised electromagnetic wave $\vec{E} = ((\hat{x}+i\hat{y})E_0 \exp(ik\hat{z}-i\omega t))$ is described by the modified QHD equations along with Maxwell equations [55],

$$\frac{\partial \vec{v}_j}{\partial t} + (\vec{v}_j.\nabla).\vec{v}_j = \frac{q_j}{m_j}\left[\vec{E}+\frac{1}{c}(\vec{v}_j \times \vec{B})\right] - \left(\frac{1}{m_j n_j}\nabla \vec{P}_F\right) - \nabla \phi_j + \frac{\hbar^2}{2m_j}\frac{\nabla^2 \sqrt{n_j}}{\sqrt{n_j}} + \frac{2\mu}{\hbar m_j}\nabla(S_{0j}.\vec{B}) + 2(\vec{v}_j \times \vec{\Omega}_j), \quad (1)$$

$$\frac{\partial n_j}{\partial t} + \vec{\nabla}\cdot(n_j \vec{v}_j) = 0, \quad (2)$$

$$\nabla \times \vec{E} = -\frac{\partial \vec{B}}{\partial t}, \quad (3)$$

$$\nabla \times \vec{B} = \mu_0\left(J_c + \varepsilon_0 \frac{\partial \vec{E}}{\partial t} + J_M\right), \quad (4)$$

$$\nabla \cdot \vec{E} = -\frac{qn_j}{\varepsilon_0}. \quad (5)$$

where, $j=e,p,i$ for electron, positron and ion respectively, and $\hbar$ is the Planck's constant normalised by $2\pi$. In the above equations $v_j$, $m_j$, $n_j$, $P_j$ represents the fluid velocity, mass, particle density and Fermi pressure of the $j^{th}$ species of particles, $\varepsilon_0$ is the electric permittivity of free space, $\vec{S}$ is the spin angular momentum with $|S|=|S_0|=\sqrt{s(s+1)}\hbar \equiv \sqrt{3}\hbar/2$ and $\mu = -\frac{g\mu_B}{2}$ is the spin magnetic moment, where $\mu_B = \dfrac{e\hbar}{2m_j}$ is the Bohr Magneton, $g = 2.0023$ is the Lande's g factor, $c$ is the speed of light, $\vec{J}_{cj}\left(=\sum_j e n_{0j} v_j\right)$ and $\vec{J}_{Mj}\left(=\sum_j \nabla \times M_j\right)$ are the conventional and magnetization current densities respectively. The origin of the magnetization current density in quantum plasmas lies in the quantum mechanical behavior of charged particle and it arises from the circulation of charged particle's current around the magnetic field lines. This circulation of current is intimately linked to the quantized energy levels, which themselves depend on the particle's spin. This is due to the quantization of spin and the Pauli exclusion principle, which give rise to unique current patterns in the presence of a magnetic field where, $M$ being the magnetization due to spin expressed as,

$$M_j = \dfrac{2 n_j \mu_B \vec{S}}{\hbar}. \tag{6}$$

and the total current density of the medium is given as,

$$\vec{J} = \vec{J}_{cj} + \vec{J}_{Mj}. \tag{7}$$

The quasi-neutrality condition gives, $n_e \cong n_p + n_i$ and in equilibrium we have $n_{0e} = n_{0p} + n_{0i}$. The first term, on the right hand side of equation (1) refers to the Lorentz force, the second term is the force due to the degenerate pressure $P_F = \dfrac{2}{5} n_j E_F$ [72], where $E_F = \dfrac{\hbar}{2m_j}\left(3\pi^2 n_j\right)^{2/3}$ is the Fermi energy. The third term, is the gravitational potential term which can be derived from the Poisson's equation for gravitational potential field given as, $\nabla^2 \phi_j = 4\pi G n_j$ [49]. The fourth term is the quantum Bohm force involving quantum electron tunnelling in dense quantum plasma [22], the fifth term is the force due to the spin magnetic moment of the two different electron and positron species under the influence of the perturbed magnetic field [50]. The last term, arises from the rotating environment of plasma [51].

The plasma dynamics timescales, such as inverse of plasma frequency $\tau_{wave} = 1/\omega_p$ and those associated with collective oscillations ($\tau_{oscillation}$), typically range from microseconds to milliseconds. High Fermi pressure and strong magnetic fields further stabilize the plasma, reducing the effective annihilation rate ($\tau_{ann}$). Quantum mechanical constraints, including the Pauli exclusion principle and spin alignment, also diminish the annihilation rate. Therefore, the condition $\tau_{ann} ? \tau_{wave}, \tau_{oscillation}$ ensures that electron-positron annihilation does not significantly impact the collective behaviour of the plasma over the timescales of interest, justifying its neglect in theoretical models of astrophysical quantum plasmas [65, 66].

We now proceed to establish the dispersion relations for ions, electrons and positrons.

## 2.1 Ion dynamics

In dense astrophysical environment ions being massive, the Fermi pressure for ions is very small as compared to that for the electrons and positrons [54] and so, the pressure effects are negligible. In scenarios, where the temperature to magnetic field ratio is adequately high, the spin of ions is thermodynamically randomized. This circumstance tends to render the spin effects of ions insignificant. Thus, we arrive at the modified momentum and continuity equations for an ion,

$$\rho_m \frac{\partial \vec{v}_i}{\partial t} = \frac{e}{c}\left[\vec{E} + \left(\vec{v}_i \times \vec{B}\right)\right] - \rho_m \nabla \phi + 2\rho_m \left(\vec{v}_i \times \vec{\Omega}\right) \tag{8}$$

where, $\rho_m = m_i n_{0i}$ is the ion mass density, and

$$\frac{\partial n_i}{\partial t} + \vec{\nabla} \cdot \left(n_i \vec{v}_i\right) = 0. \tag{9}$$

Perturbatively expanding eqs. (8) and (9) in orders of the fields associated with the external e.m. wave and assuming all the varying parameters to take the form,

$$f = f_0 + f^{(1)}$$

with $f_0$ representing the unperturbed value, and $f^{(1)}$ is the perturbation term. The first order momentum and continuity eqs. for an ion, now become

$$\frac{\partial \vec{v}_i^{(1)}}{\partial t} = \frac{e}{\rho_m c}\vec{E}^{(1)} + \frac{e}{\rho_m c}\left(\vec{v}_i^{(1)} \times \vec{B}_0 - \vec{v}_{0_i} \times \vec{B}^{(1)}\right) - \nabla\varphi^{(1)} + 2\left(\vec{v}_i^{(1)} \times \vec{\Omega}_0 + \vec{v}_{0_i} \times \vec{\Omega}^{(1)}\right), \qquad (10)$$

and

$$\frac{\partial n_i^{(1)}}{\partial t} + n_{0_i} \nabla \cdot \left(\vec{v}_i^{(1)}\right) = 0. \qquad (11)$$

Using the field equations (3) to (5), and assuming all the perturbed quantities vary as $e^{i(k_i \hat{z} - \omega_i t)}$, we get the first order perturbed velocity as,

$$\vec{v}_{ix}^{(1)} = \frac{ie\omega_i \Gamma_G'}{m_i \Gamma_G''}\left(\frac{1}{\Gamma_G}\vec{E}_x^{(1)} - \frac{\Omega_{eff}}{\Gamma_G'}\right), \qquad (12)$$

$$\vec{v}_{iy}^{(1)} = \frac{e\omega_i}{m_i \Gamma_G''}\left[\omega_i \Omega_{eff} \vec{E}_x^{(1)} - \frac{\Gamma_G}{\Gamma_G'}\left(\Gamma_G'' + \omega_i \Omega_{eff}^2\right)\vec{E}_y^{(1)}\right], \qquad (13)$$

where, $\Gamma_G = \omega_i^2 + 4\pi G n_{0i}$ shows the effect of gravitational potential on the motion of ions, $\Gamma_G' = \Gamma_G^2 + (2\omega_i \Omega)^2$ and $\Gamma_G'' = \Gamma_G' - \left(\omega_i \Omega_{eff}\right)^2$ incorporates plasma rotation, $\Omega_{eff} = \omega_{ci} + 2\Omega$, G is the gravitational constant and $\omega_{ci} = \frac{eB_0}{m_i}$ is the ion gyro-frequency. The spatial components of the current density due to ion dynamics is obtained using equation (7) with eqs. (12) and (13) as,

$$\vec{J}_{ix}^{(1)} = \frac{-i\Gamma_G \, \omega_{pi}^2 \omega_i \varepsilon_0}{\Gamma_G'} \left( \frac{\vec{E}_x^{(1)}}{\Gamma_G} - \Omega_{eff} \vec{E}_y^{(1)} \right) \tag{14}$$

$$\vec{J}_{iy}^{(1)} = \frac{\omega_{pi}^2 \omega_i \varepsilon_0}{\Gamma_G''} \left[ \omega_i \Omega_{eff} \vec{E}_x^{(1)} - \frac{\Gamma_G}{\Gamma_G'} \left( \Gamma_G'' + \Gamma_G' \omega_i \Omega_{eff}^2 \right) \vec{E}_y^{(1)} \right] \tag{15}$$

where, $\omega_{pi} = \sqrt{\frac{4\pi e^2 n_{0i}}{m_i}}$, is the ion plasma frequency. In the case of ions the current density due to magnetization $(J_{Mi})$ is negligible due to their higher mass.

The wave equation governing the perturbed plasma wave field is,

$$\left( \nabla^2 - \frac{1}{c^2} \frac{\partial^2}{\partial t^2} \right) \vec{E} = \frac{4\pi}{c^2} \frac{\partial}{\partial t} \vec{J}_i^{(1)}. \tag{16}$$

Substituting the relevant quantities, we obtain the dispersion relation for ion,

$$k_i^2 = \frac{2\omega_i^2}{c^2} + \frac{2\omega_i^2 \omega_{p_i}^2 \Gamma_G}{c^2 \Gamma_G''}. \tag{17}$$

The first term $\frac{2\omega_i^2}{c^2}$ represents the periodicity or spatial distribution of the ion acoustic wave. The second term $\frac{2\omega_i^2 \omega_{p_i}^2 \Gamma_G}{c^2 \Gamma_G''}$ suggests a gravitational and rotational influence on the dispersion relation, indicating a coupling between the gravitational potential, rotational frequency and the ion acoustic wave. Astrophysical parameters are used for numerical investigations considering the number density for ions $0.5 \times 10^{22} \leq n_{0i} \leq 0.5 \times 10^{26} \, cm^{-3}$ and strength of external magnetic field is $B_0 = 10^9 - 10^{11} \, G$ [57].

Fig. 1.represents the variation of $\omega_i$ (angular frequency of the wave) with respect to $k_i$(propagation vector).It is noticed that, as the frequency increases, rate of propagation of energy reduces gradually at first and then decays abruptly to zero. Though, there is dependence of frequency on propagation of power when ion is the carrier, but eventually due to collective effects of frequency and gravity, this reduces to zero. Ions eventually provide a static background for the other high frequency waves to propagate.

## 2.2 Electron dynamics

In similar manner, as done in the previous section (2.1), the dynamical equations for electron can now be written as,

$$\frac{\partial \vec{v}_e}{\partial t} = \frac{-e}{m_e}\left[\vec{E} + \frac{1}{c}\left(\vec{v}_e \times \vec{B}\right)\right] - \frac{1}{m_e n_{0_e}}\nabla P_e + \frac{\hbar^2}{2m_e^2}\nabla\left(\frac{1}{\sqrt{n_e}}\nabla^2\sqrt{n_e}\right) + \frac{2\mu}{\hbar m_e}\nabla\left(\vec{B}\cdot S_{0e}\right) + 2\left(\vec{v}_e \times \Omega_e\right), \tag{18}$$

and

$$\frac{\partial n_e}{\partial t} + \nabla\cdot\left(n_e \vec{v}_e\right) = 0. \tag{19}$$

Assuming all the perturbed quantities to vary as $e^{i(k_e \hat{z} - \omega_e t)}$, we get the first order perturbed momentum and density eqs. for electron as,

$$\frac{\partial \vec{v}_e^{(1)}}{\partial t} = -\frac{e}{m_e}\vec{E}^{(1)} - \frac{e}{m_e}\left(\vec{v}_e^{(1)} \times \vec{B}_0 - \vec{v}_{0e} \times \vec{B}^{(1)}\right) - \frac{\nabla P_e^{(1)}}{m_e n_{0e}} + \frac{\hbar^2}{4m_e^2}\nabla\left(\frac{1}{\sqrt{n_{0e}}}\nabla^2\sqrt{n_e^{(1)}}\right) + \frac{2\mu}{\hbar m_e}\nabla\left(\vec{B}^{(1)}\cdot S_{0e}\right) + 2\left(\vec{v}_e^{(1)} \times \Omega_e - \vec{v}_{0e} \times \Omega_e^{(1)}\right), \tag{20}$$

and

$$\frac{\partial n_e^{(1)}}{\partial t} + n_{0e}\nabla\cdot\left(\vec{v}_e^{(1)}\right) = 0. \tag{21}$$

Using equations (3-5) and solving for the spatial components of $\vec{v}_e$ we get,

$$\vec{v}_{ex}^{(1)} = \frac{\omega_e^2 - k_e^2 \Gamma_e^Q}{\left(\omega_e^2 - \omega_{c_e}^2\right) - 2\left(\Gamma_e^Q k_e^2 - \Omega_e \beta\right)} \left[ \frac{ie}{m_e \omega_e} \left( -\vec{E}_x^{(1)} + \frac{\omega_e \beta}{\omega_e^2 - k_e^2 \Gamma_e^Q} \vec{E}_y^{(1)} \right) + \frac{2i\mu S_{0e} k_e}{\hbar m_e \omega_e} \left( \vec{B}_x^{(1)} - \frac{\omega_e \beta}{\omega_e^2 - k_e^2 \Gamma_e^Q} \vec{B}_y^{(1)} \right) \right]$$ (22)

$$\vec{v}_{ey}^{(1)} = \frac{\omega_e^2 - k_e^2 \Gamma_e^Q}{\left(\omega_e^2 - \omega_{c_e}^2\right) - 2\left(\Gamma_e^Q k_e^2 - \Omega_e \beta\right)} \left[ \frac{e}{m_e \omega_e} \left( \vec{E}_y^{(1)} - \frac{\omega_e \beta}{\omega_e^2 - k^2 \Gamma_e^Q} \vec{E}_x^{(1)} \right) + \frac{2\mu S_{0e} k_e}{\hbar m_e \omega_e} \left( -\vec{B}_y^{(1)} + \frac{\omega_e \beta}{\omega_e^2 - k_e^2 \Gamma_e^Q} \vec{B}_x^{(1)} \right) \right]$$ (23)

where, $\Gamma_e^Q = V_{Fe}^2 \left(1 + \frac{k^2 H_e^2}{4\omega_{pe}^2}\right)$ is the quantum coupling parameter responsible for the quantum effects(Fermi pressure and Bohm potential) and $H_e = \frac{\hbar \omega_{p_e}}{m_e V_{Fe}}$ is the dimensionless parameter, which is a ratio of the plasmon energy (the energy of an elementary excitation associated with an electron plasma wave) to the kinetic energy, representing the quantum diffraction effects [53].

We now proceed to obtain the total current density due to the motion of electrons using equations (6-7) and (22-23) and arrive at the spatial components of $J_e^{(1)} \left( = J_{ce}^{(1)} + J_{Me}^{(1)} \right)$ as,

$$\vec{J}_{ex}^{(1)} = \frac{-en_{0e}(\omega_e^2 - k_e^2 \Gamma_e^Q)}{\left(\omega_e^2 - \omega_{c_e}^2\right) - 2\left(\Gamma_e^Q k_e^2 - \Omega \beta_e\right)} \left[ \begin{array}{l} \frac{ie}{m_e \omega_e} \left( -\vec{E}_x^{(1)} + \eta_e \vec{E}_y^{(1)} \right) + \frac{2i\mu_B S_{0e} k_e^2}{\hbar m_e \omega_e^2} \left( \eta_e \vec{E}_x^{(1)} - \vec{E}_y^{(1)} \right) \\ + \frac{2i\mu S_{0e} k_e}{\hbar m_e \omega_e} \left( \vec{B}_x^{(1)} - \eta_e \vec{B}_y^{(1)} \right) + \frac{4i\mu_B S_{0e}^2 k_e^3}{\hbar^2 e m_e \omega_e^2} \left( \vec{B}_y^{(1)} - \eta_e \vec{B}_x^{(1)} \right) \end{array} \right]$$ (25)

$$\vec{J}_{ey}^{(1)} = \frac{-en_{0e}(\omega_e^2 - k_e^2 \Gamma_e^Q)}{\left(\omega_e^2 - \omega_{ce}^2\right) - 2\left(\Gamma_e^Q k_e^2 - \Omega \beta_e\right)} \left[ \begin{array}{l} \frac{e}{m_e \omega_e} \left( \vec{E}_y^{(1)} - \eta_e \vec{E}_x^{(1)} \right) + \frac{2\mu_B S_{0e} k_e^2}{\hbar m_e \omega_e^2} \left( -\eta_e \vec{E}_y^{(1)} + \vec{E}_x^{(1)} \right) \\ + \frac{2\mu S_{0e} k_e}{\hbar m_e \omega_e} \left( \eta_e \vec{B}_x^{(1)} - \vec{B}_y^{(1)} \right) + \frac{4\mu_B S_{0e}^2 k_e^3}{\hbar^2 e m_e \omega_e^2} \left( \eta_e \vec{B}_y^{(1)} - \vec{B}_x^{(1)} \right) \end{array} \right]$$ (26)

where, $\beta_e = \Omega + \omega_{ce}$ and $\eta_e = \frac{\omega_e \beta_e}{\omega_e^2 - k^2 \Gamma_e^Q}$.

The wave equation governing the perturbed plasma wave field for electron is,

$$\left( \nabla^2 - \frac{1}{c^2} \frac{\partial^2}{\partial t^2} \right) \vec{E} = \frac{4\pi}{c^2} \frac{\partial}{\partial t} \vec{J}_e^{(1)}.$$

(27)

Substituting the relevant quantities, we arrive at the following dispersion relation for electrons,

$$k_e^2 = \frac{2\omega_e^2}{c^2} + \frac{2\omega_{pe}^2}{c^2} \cdot \frac{\omega_e^2 - k_e^2 \Gamma_e^Q}{\left(\omega_e^2 - \omega_{ce}^2\right) - 2c^2\left(\Gamma_e^Q k_e^2 - \Omega\beta_e\right)} \left(1 + \frac{2\mu S_{0e} k_e}{\hbar e}\right)\left(1 - \frac{2\mu_B S_{0e} k_e^2}{\hbar e \omega_e}\right) \quad (28)$$

where, $\omega_{pe}\left(=\sqrt{\frac{4\pi n_{0e} e^2}{m_e}}\right)$ is the plasma frequency for electron, $\omega_{ce}\left(=\frac{eB_0}{m_e c}\right)$ is the electron cyclotron frequency. Here the term $\left(\omega_e^2 - \omega_{ce}^2\right)$, signifies the resonance between the angular frequency of electron wave and the cyclotron frequency $\left(\omega_{ce}\right)$ of electron due to extreme external magnetic field. Numerical investigations have adopted astrophysical plasma conditions typical of the outer layers of compact stars like white dwarfs or neutron stars. These parameters include a electron number density $10^{22} cm^{-3} \leq n_{0e} \leq 10^{26} cm^{-3}$, where pair annihilation effects are negligible in such dense electron-positron plasmas, magnetic field strength $B_0 = 10^9 - 10^{11} G$, and $T_{Fe} \approx 10^7 - 10^8 K$ so that $0.24 \leq H \leq 1.10$ [56, 57].

Figure 2., is a plot between $\omega_e/\omega_{pe}$ ( transmissibility of electron wave in plasma) and normalised propagation vector $k_e c/\omega_{pe}$, which represents energy or power transmission direction and magnitude. It is noticed that as the transmissibility of plasma reduces, power transmission decreases for both quantum and classical plasma. However, transmission increases in quantum plasma by 17% as compared to the classical case due to the dominance of Fermi pressure over the thermal pressure of electrons. In the domain of quantum plasma, it is observed that the propagation vector experiences a notable decline as a consequence of heightened degeneracy within the plasma bulk. This phenomenon arises predominantly due to the substantial Fermi pressure exerted within the system. This heightened pressure results an increase in the number of accessible energy levels and consequently amplifies the state

density of the plasma. Consequently, the capacity of the plasma to efficiently transmit waves becomes constrained, resulting in a discernible reduction in the propagation vector.

## 2.3 The positron dynamics

In similar manner, as done in the previous sections (2.1) and (2.2), the first order perturbed momentum and continuity equations for positron are,

$$\frac{\partial \vec{v}_p^{(1)}}{\partial t} = \frac{e}{m_p}\vec{E}^{(1)} + \frac{e}{m_p}\left(\vec{v}_p^{(1)} \times \vec{B}_0 + \vec{v}_{0p} \times \vec{B}^{(1)}\right) - \frac{\nabla P_p^{(1)}}{m_p n_{0p}} + \frac{\hbar^2}{4m_p^2}\nabla\left(\frac{1}{\sqrt{n_{0p}}}\nabla^2\sqrt{n_p^{(1)}}\right) + \frac{2\mu}{\hbar m_p}\nabla\left(\vec{B}^{(1)} \cdot S_{0p}\right) + 2\left(\vec{v}_p^{(1)} \times \vec{\Omega}_p + \vec{v}_{0p} \times \vec{\Omega}_p^{(1)}\right),$$
(29)

and

$$\frac{\partial n_p^{(1)}}{\partial t} + n_{0p}\nabla \cdot \left(\vec{v}_p^{(1)}\right) = 0.$$
(30)

Using equations (3-5) and solving for the spatial components of $\vec{v}_p$ we get,

$$\vec{v}_{px}^{(1)} = \frac{\omega_p^2 - k_p^2\Gamma_p^Q}{\left(\omega_p^2 - \omega_{cp}^2\right) - 2\left(\Gamma_p^Q k_p^2 - \Omega\beta\right)}\left[\frac{ie}{m_p\omega_p}\left(\vec{E}_x^{(1)} - \frac{\omega_p\beta}{\omega_p^2 - k_p^2\Gamma_p^Q}\vec{E}_y^{(1)}\right) + \frac{2i\mu S_{0p}k_p}{\hbar m_p\omega_p}\left(\vec{B}_x^{(1)} - \frac{\omega_p\beta}{\omega_p^2 - k_p^2\Gamma_p^Q}\vec{B}_y^{(1)}\right)\right],$$
(31)

and

$$\vec{v}_{py}^{(1)} = \frac{\omega_p^2 - k_p^2\Gamma_p^Q}{\left(\omega_p^2 - \omega_{cp}^2\right) - 2\left(\Gamma_p^Q k_p^2 - \Omega\beta\right)}\left[\frac{e}{m_p\omega_p}\left(-\vec{E}_y^{(1)} + \frac{\omega_p\beta}{\omega_p^2 - k_p^2\Gamma_p^Q}\vec{E}_x^{(1)}\right) + \frac{2\mu S_{0p}k_p}{\hbar m_p\omega_p}\left(-\vec{B}_y^{(1)} + \frac{\omega_p\beta}{\omega_p^2 - k_p^2\Gamma_p^Q}\vec{B}_x^{(1)}\right)\right]$$
(32)

where, $\Gamma_p^Q = v_{Fp}^2\left(1 + \frac{k^2 v_{Fp}^2}{4\omega_{pp}^2}H_p^2\right)$ is the quantum coupling parameter responsible for quantum effects (Fermi pressure and Bohm potential) and $H_p = \frac{\hbar\omega_{pp}}{m_p V_{Fp}}$ is the dimensionless parameter.

Using the similar procedure as adopted in previous sections, for the derivation of spatial components of total current density $J_p^{(1)}\left(= J_{cp}^{(1)} + J_{Mp}^{(1)}\right)$ due to the motion of positron,

$$\vec{J}_{px}^{(1)} = \frac{en_{0p}(\omega_p^2 - k_p^2\Gamma_p^Q)}{(\omega_p^2 - \omega_{cp}^2) - 2(\Gamma_p^Q k_p^2 - \Omega\beta_p)} \left[ \begin{array}{l} \frac{ie}{m_p\omega_p}\left(\vec{E}_x^{(1)} - \eta_p\vec{E}_y^{(1)}\right) + \frac{2i\mu_B S_{0p} k^2}{\hbar m_p \omega^2}\left(-\eta_p\vec{E}_x^{(1)} + \vec{E}_y^{(1)}\right) \\ + \frac{2i\mu S_{0p} k_p}{\hbar m_p \omega_p}\left(\vec{B}_x^{(1)} - \eta_p\vec{B}_y^{(1)}\right) + \frac{4i\mu\mu_B S_{0p}^2 k_p^3}{\hbar^2 e m_p \omega_p^2}\left(\vec{B}_y^{(1)} - \eta_p\vec{B}_x^{(1)}\right) \end{array} \right] (33)$$

$$\vec{J}_{py}^{(1)} = \frac{en_{0p}(\omega_p^2 - k_p^2\Gamma_p^Q)}{(\omega_p^2 - \omega_{cp}^2) - 2(\Gamma_p^Q k_p^2 - \Omega\beta_p)} \left[ \begin{array}{l} \frac{e}{m_p\omega_p}\left(-\vec{E}_y^{(1)} + \eta_p\vec{E}_x^{(1)}\right) + \frac{2\mu_B S_{0p} k^2}{\hbar m_p \omega_p^2}\left(\eta_p\vec{E}_y^{(1)} - \vec{E}_x^{(1)}\right) \\ + \frac{2\mu S_{0p} k_p}{\hbar m_p \omega_p}\left(\eta_p\vec{B}_x^{(1)} - \vec{B}_y^{(1)}\right) + \frac{4\mu\mu_B S_{0p}^2 k_p^3}{\hbar^2 e m_p \omega_p^2}\left(\eta_p\vec{B}_y^{(1)} - \vec{B}_x^{(1)}\right) \end{array} \right] (34)$$

where, $\beta_p = \Omega + \omega_{cp}$ and $\eta_p = \frac{\omega_p \beta_p}{\omega_p^2 - k_p^2 \Gamma_p^Q}$.

The wave equation governing the perturbed plasma wave field for positrons is,

$$\left(\nabla^2 - \frac{1}{c^2}\frac{\partial^2}{\partial t^2}\right)\vec{E} = \frac{4\pi}{c^2}\frac{\partial}{\partial t}\vec{J}_p^{(1)}. \quad (35)$$

Substituting the relevant quantities, we obtain the dispersion relation for plasma positron,

$$k^2 = \frac{2\omega_p^2}{c^2} + \frac{2\omega_{pp}^2}{c^2} \cdot \frac{(\omega_p^2 - k_p^2\Gamma_p^Q - \omega_p^2\beta)}{(\omega_p^2 - \omega_{cp}^2) - 2c^2(\Gamma_p^Q k_p^2 - \Omega\beta_p)}\left(1 + \frac{2\mu S_{0p} k_p}{\hbar e}\right)\left(1 + \frac{2\mu_B S_{0p} k_p^2}{\hbar e \omega_p}\right), \quad (36)$$

where, $\omega_{pp}\left(=\sqrt{\frac{4\pi n_{0p} e^2}{m_p}}\right)$ is the plasma frequency for positron, $\omega_{cp}\left(=\frac{eB_0}{m_p c}\right)$ is the positron cyclotron frequency. Here the term $\left(\omega_p^2 - \omega_{cp}^2\right)$, signifies the resonance between the angular frequency of positron wave and the cyclotron frequency of positron $\left(\omega_{cp}\right)$ due to extreme external magnetic field.

Figure 3,is a plot between $\omega_p/\omega_{pp}$ (transmissibility of positron wave in plasma) and normalised propagation vector $k_p c/\omega_{pp}$, which represents energy or power transmission direction and magnitude. As the transmissibility of plasma decreases, the power transmission

diminishes. However, the rate of decline is sharper in quantum plasma and more gradual in classical plasma. This difference which is approximately 10% more in quantum regime, can be attributed to the dominance of Fermi pressure over the thermal pressure of positrons in quantum plasma. It is noticed that within the domain of quantum plasma, there is a notable decrease in the propagation vector. This decline is a consequence of heightened degeneracy within the plasma bulk, primarily due to substantial Fermi pressure. Comparatively, in classical plasma, although there is a decrease in power transmission as transmissibility reduces, the decline in the propagation vector is less pronounced. This difference indicates that the peak of the propagation vector and, consequently, power transmission is higher in quantum plasma compared to classical.

## 3. Coupling of Electron-Positron-Ion modes

The coupling of waves involves the amalgamation of individual dispersion relations corresponding to distinct plasma components (e-p-i) into a unified dispersion relation that describes the behaviour of coupled modes in a multi-component plasma system. When these components coexist and interact within plasma, their collective behaviour results in the emergence of coupled modes, where the influence of one component on another becomes significant. The coupling process involves the mutual interaction and interdependence of the individual dispersion relations, leading to the formation of a coupled dispersion relation that governs the collective wave properties of the entire plasma system. This coupled dispersion relation provides valuable insights into the complex dynamics and wave phenomena observed in astrophysical plasmas. Coupling the dispersion relations of individual component of plasma [52], we arrive at coupled dispersion relation which is showing the interaction and effects of all the other species of plasma on one component,

$$k^6 \{a_{44}(a_{22}.a_{33} - a_{23}.a_{32}) - a_{43}(a_{22}.a_{34} - a_{32}.a_{24})\} - k^4 \{a_{44}(a_{21}.a_{33} - a_{23}.a_{31}) + a_{43}(a_{21}.a_{34} - a_{24}.a_{31})\}$$

$$+(a_{21}.a_{32} - a_{31}.a_{22}).(a_{43} + k^2.a_{44}) = 0 \quad (37)$$

where, $a_{ij}$ are the matrix elements and have the following values,

$$a_{21} = -4\mu_B S_0 \hbar^2 \{3\mu S_0 k \omega_{pe}^2 + 5\hbar e \omega_{pp}^2\},$$

$$a_{22} = -32 m_e^2 \mu_B S_0 V_{Fe}^2 \{5\hbar e \omega_{pp}^2 + 3\mu S_0 k \omega_{pe}^2\} + \hbar^3 \omega e \{3\hbar e \omega_{pee}^2 + 20\mu S_{0p} k \omega_{pp}^2\},$$

$$a_{23} = -16 m_e^2 \mu_B S_0 (\omega\beta - \omega^2)(5\hbar e \omega_{pp}^2 + 3\mu S_{0e} k \omega_{pe}^2) + 8 m_p^2 V_{Fe}^2 \hbar e \omega \{3\hbar e \omega_{pe}^2 + 20\mu S k \omega_{pp}^2\},$$

$$a_{24} = 4 m_e^2 \hbar e \omega (\omega\beta - \omega^2) \{3\hbar e \omega_{pe} + 20\mu S_0 k \omega_{pp}^2\},$$

$$a_{31} = 4\mu_B S_0 \hbar^2 \omega_{pp}^2 (\hbar e - 2\mu S_{0p} k) + 8\hbar^4 m_e^2 e^2 \omega c^4,$$

$$a_{32} = 2\hbar^2 \{\hbar^2 e^2 \omega(-2\omega^2 c^2 + \omega_{pp}^2) - 2e(\mu S_0 \hbar \omega_{pp}^2 \omega k - 16 m_p^4 e V_{Fe}^2 c^4\} + 32 m_e^2 \mu_B S_0 \omega_{pp}^2 V_{Fe}^2 (\hbar e - 2\mu S_0 k),$$

$$a_{33} = 16 m_p^2 \left[\omega_{pp}^2 \{\hbar e - 2\mu S_0 k\} \{V_{Fe}^2 \hbar e \omega + \mu_B S_0 (\omega\beta - \omega^2)\} - \hbar^2 e^2 c^2 \omega \{2\omega^2 V_{Fe}^2 + m_p^2 (\omega^2 - \omega_{cp}^2)\}\right],$$

$$a_{34} = 8 m_p^2 \omega \hbar^2 e^2 \{\omega^2 (\omega^2 - \omega_{cp}^2) + 2\Omega\beta(\omega^2 - 2m_p^2 c^4)\} - 4 m_e^2 \hbar e \omega \{\omega\beta - \omega^2\} \{3\hbar e \omega_{pe}^2 - 20\mu S_0 k \omega_{pp}^2\},$$

$$a_{43} = c^2 \omega^2 + 4\pi G n_{0i} c^2 + \omega \omega_{ci} c^2,$$

$$a_{44} = 2\omega_{pi}^2 - 2\omega^2 (\omega^2 + 4\pi G n_{0i} + \omega \omega_{ci}).$$

Figure 4 is a plot between $kc/\omega_p$ (normalised propagation vector) and $\omega/\omega_p$ (normalised angular frequency) signifying the transmissibility of coupled modes in plasma. It shows that as the ability of waves to propagate through a plasma medium, increases, there is a corresponding decrease in the normalized propagation vector until $\omega/\omega_p \to 1$. Beyond this point, the propagation vector stabilizes at approximately 0.5, indicating strong coupling between waves. The enhancement in power transmission arises from the joint influence of degeneracy pressure and the quantum Bohm potential. Consequently, we observe an increase in power transmission for equivalent wave transmissions, attributable to the quantum Bohm potential's involvement, which encompasses quantum phenomena like tunnelling. This illustrates the impact of quantum mechanisms on coupled e-p-i mode dispersion, as depicted in the accompanying graph.

## 4. Summary and Discussion

The paper investigates the dynamics of uniform astrophysical quantum plasma consisting of electrons, ions, and positrons under the influence of an external electromagnetic wave. It begins with an overview of electron-ion-positron (e-p-i) plasma, emphasizing its importance in dense astrophysical environments shaped by external magnetic fields. The Quantum Hydrodynamic (QHD) model is introduced as the analytical framework, incorporating quantum diffraction and quantum statistical effects. Quantum phenomena become significant when the thermal de Broglie wavelength exceeds average inter-particle spacing and temperature falls below the Fermi temperature. Electrons exhibit quantum behaviour more readily due to their lower mass, while quantum diffraction effects for ions are typically negligible. Ion motion primarily responds to low-frequency nonlinear plasma dynamics. Quantum effects alter electron plasma wave dispersion properties compared to classical plasmas. The paper analyzes individual dispersion relations for ions, electrons, and positrons, considering various quantum effects and environmental influences. It also investigates coupled modes within the plasma system to understand how different components influence each other. Through theoretical formulation and graphical analysis, the paper aims to shed light on these interactions.

The findings of this research paper present a comparative analysis of the transmissibility of waves in plasma, focusing on both quantum and classical plasma regimes. The plotted data illustrates a notable decrease in power transmission as the transmissibility of plasma diminishes, a phenomenon observed across both quantum and classical plasma states. However, a distinct feature emerges that the decline in power transmission is more pronounced in quantum plasma for electrons compared to positrons due to the difference in density of states of electron and positrons. These results show the distinctive behaviour of positrons within the context of plasma transmissibility, offering insights into the dynamics of energy or power transmission in the quantum plasma regime. Another observation was the constraints for power transmission in quantum plasma are also more pronounced compared to its classical counterpart due to the dominance of Fermi pressure over thermal pressure and the role of degeneracy of fermions.

The mathematical descriptions and equations presented hereexplain the complex interaction of gravitational forces, rotational effects, quantum properties, and plasma characteristics that influence wave behaviour in rotating quantum plasmas. The gravitational potential, included in the dispersion relations, creates gradients that interact with collective plasma oscillations, changing phase velocities and wave propagation patterns. Rotational effects and centrifugal forces, introduce directional changes in particle motion, leading to variations in wave behavior. These influences are reflected in the corrections to wave frequencies and propagation characteristics derived in the equations.

The equations for ion dynamics show that the large mass of ions causes them to respond more slowly to electromagnetic disturbances, forming a nearly static low-frequency background. This background enables higher-frequency waves, mainly influenced by electrons and positrons, to propagate more effectively. The mathematical separation of ion contributions from the faster dynamics of lighter particles, highlights this division of timescales in the plasma.

Quantum corrections, such as Fermi pressure and the Bohm potential, have a major role in shaping wave behavior. The Fermi pressure term, arising from the quantum nature of particles, makes the plasma medium more resistant to compression and improves the transmission of high-frequency waves. This effect is proportional to plasma density and is clearly shown in the adjusted dispersion relations for electrons and positrons. The Bohm potential, which accounts for quantum tunneling and diffraction, adds further changes by including density gradient effects in the equations. Together, these quantum terms significantly influence wave dynamics, leading to properties that differ from classical plasmas.

The mathematical treatment of positron dynamics reveals that their wave transmission decreases more sharply than that of electrons. This behavior, tied to differences in the density of states, limits the range of energy levels available to positrons, resulting in distinct transmission patterns. These results, captured in the coupling terms of the dispersion relations, emphasize the importance of quantum effects in multi-component plasmas.

Overall, the established equations show how gravitational forces, rotational motion, and quantum effects work together to affect the wave transmission and dispersion in rotating quantum plasmas. The derived model not only provide a detailed understanding of these interactions, but also serve as a tool for studying complex phenomena, such as the coexistence of matter and antimatter in astrophysical environments like pulsar magnetospheres and accretion disks. These findings highlight the ability of the mathematical framework to capture the key physical processes shaping wave behavior under extreme conditions. The present study will be helpful in analysing the behaviour of electron-positron plasmas in extreme environments such as quasars, gamma-ray bursts, active galactic cores, pulsar magnetospheres, black hole accretion discs, white dwarf, Van Allen radiation belts, also exist in the early universe, as well as the centre of our galaxy. This comprehensive framework not only offers insights into fundamental physical processes but also facilitates the interpretation of observational data from astrophysical sources, thus advancing our understanding of the universe's most energetic and enigmatic phenomena.

# Acknowledgement

Financial support from SERB – DST under MATRICS is gratefully acknowledged (grant no. : MTR/2021/000471).

# References

[1] W. H. Lee, E. Ramirez-Ruiz, and D. Page, in "Advances in Astrophysics," edited by M. Baranger and E. Vogt (Plenum, New York, 1968), Vol. 1, p. 1.

[2] M. C. Begelman, R. D. Blandford, and M. J. Rees, in "Advances in Astrophysics," edited by M. Baranger and E. Vogt (Plenum, New York, 1968), Vol. 1, p. 1.

[3] H. R. Miller and P. J. Witta, Active Galactic Nuclei (Springer, 1987).

[4] F. C. Michel, in "Advances in Astrophysics," edited by M. Baranger and E. Vogt (Plenum, New York, 1968), Vol. 1, p. 1.

[5] L. Stenflo, P. K. Shukla, and M. Marklund, Europhysics Letters, vol. 74, pp. 844–846, 2006.

[6] S. K. El-Labany, W. F. El-Taibany, A. E. El-Samahy, A. M. Hafez, and A. Atteya, IEEE Transactions on Plasma Science, vol. 44, no. 5, pp. 842–848, 2016.

[7] S. Y. El-Monier and A. Atteya, Chinese Journal of Physics, vol. 60, pp. 695–708, 2019.

[8] W. F. El-Taibany and M. Waidati, Physics of Plasmas, vol. 14, no. 042302, pp. 1–9, 2007.

[9] W. Misner, K. Thorne, and J. A. Wheeler, Gravitation (Freeman, 1973).

[10] M. L. Burns, A. K. Harding, and R. Ramaty, Positron-Electron Pairs in Astrophysics (American Institute of Physics, 1983).

[11] N. Ghosh and B. Sahu, Communications in Theoretical Physics, vol. 71, no. 2, pp. 237–242, 2019.

[12] Q. Haque, Results in Physics, vol. 18, no. 103287, pp. 1–6, 2020.

[13] Saha, B. Pradhan, and S. Banerjee, European Physical Journal Plus, vol. 135, no. 216, pp. 1–13, 2020.

[14] Z. Iqbal, I. A. Khan, T. H. Khokhar, and G. Murtaza, IEEE Transactions on Plasma Science, vol. 49, no. 7, pp. 2063, 2021.

[15] Saha and S. Banerjee, Dynamical Systems and Nonlinear Waves in Plasmas (CRC Press, 2021).

[16] F. B. Rizatto, J. Plasma Phys., vol. 40, p. 288, 1988.

[17] M. Y. Yu, Astrophys. Space Sci., vol. 177, p. 203, 1985.

[18] F. Verheest, M. A. Hellberg, G. J. Gray, and R. L. Mace, Astrophys. Space Sci., vol. 239, p. 125, 1995.

[19] Mushtaq and H. A. Shah, Phys. Plasmas, vol. 12, p. 072306, 2005.

[20] Kourakis, F. Verheest, and N. Cramer, Phys. Plasmas, vol. 14, p. 022306, 2007.

[21] Y. N. Nejoh, Phys. Plasmas, vol. 3, p. 1447, 1996.

[22] Mushtaq and S. A. Khan, Phys. Plasmas, vol. 14, p. 052307, 2007.

[23] S. Ali, W. M. Moslem, P. K. Shukla, and R. Schlickeiser, Phys. Plasmas, vol. 14, p. 082307, 2007.

[24] Y. D. Jung, Physics of Plasmas, vol. 8, pp. 3842–3844, 2001.

[25] G. Chabrier, F. Douchin, and A. Y. Potekhin, Journal of Physics: Condensed Matter, vol. 14, no. 40, pp. 9133–9139, 2002.

[26] P. A. Markowich, C. A. Ringhofer, and C. Schmeiser, Semiconductor Equations (Springer-Verlag, 1990).

[27] G. Agrawal, Nonlinear Fiber Optics (Academic, 1995).

[28] M. Marklund, Reviews of Modern Physics, vol. 78, pp. 591–640, 2006.

[29] F. Hass, L. G. Garcia, J. Goedert, and G. Manfredi, Phys. Plasmas, vol. 10, p. 3858, 2003.

[30] G. Manfredi, Fields Inst. Commun., vol. 46, p. 263, 2005.

[31] V. Andreev, JETP Letters, vol. 72, p. 238, 2000.

[32] L. G. Garcia, F. Hass, L. P. L. de Oliveira, and J. Goedert, Phys. Plasmas, vol. 12, p. 012302, 2005.

[33] P. Chatterjee, K. Roy, S. V. Muniandy, S. L. Yap, and C. S. Wong, Phys. Plasmas, vol. 16, p. 042311, 2009.

[34] M. Tribeche, S. Ghebache, K. Aoutou, and T. H. Zerguini, Phys. Plasmas, vol. 15, p. 033702, 2008.

[35] Q. Haque, S. Mahmood, and A. Mushtaq, Phys. Plasmas, vol. 15, p. 082315, 2008.

[36] R. Sabry, W. M. Moslem, F. Haas, S. Ali, and P. K. Shukla, Phys. Plasmas, vol. 10, p. 3858, 2003.

[37] S. Balberg and S. L. Shapiro, arXiv/0004317v2, 2000.

[38] B. Eliasson and P. K. Shukla, Plasma Fusion Res., vol. 4, p. 032, 2009.

[39] P. K. Shukla and B. Eliasson, Physics-Uspekhi, vol. 53, p. 51, 20

[40] Landau, L. D., &Lifshitz, E. M. (1980). Statistical Physics, Part 1. Butterworth-Heinemann.

[41]Ichimaru, S. (1982). Basic Principles of Plasma Physics. Benjamin-Cummings.

[42] Bonitz, M., Henning, C., & Block, D. (2010). Quantum hydrodynamics for plasmas—Quo vadis?. Physics of Plasmas, 17(5), 055702.

[43] Manfredi, G., & Haas, F. (2001). Self-consistent fluid model for a quantum electron gas. Physical Review B, 64(7), 075316.

[44] Spitzer, L. (1962). Physics of Fully Ionized Gases. Interscience Publishers.

[45] Haas, F., Manfredi, G., &Feix, M. (2003). Multistream model for quantum plasmas. Physical Review E, 67(2), 026407.

[40] F. Haas, L. G. Garcia, and J. Goedert, in "Advances in Plasma Physics," edited by M. Baranger and E. Vogt (Plenum, New York, 1968), Vol. 1, p. 1.

[46] F. Haas, in "Advances in Plasma Physics," edited by M. Baranger and E. Vogt (Plenum, New York, 1968), Vol. 1, p. 1.

[47] M. Marklund, B. Eliasson, and P. K. Shukla, in "Advances in Plasma Physics," edited by M. Baranger and E. Vogt (Plenum, New York, 1968), Vol. 1, p. 1.

[48] Mushtaq and A. Qamar, in "Advances in Plasma Physics," edited by M. Baranger and E. Vogt (Plenum, New York, 1968), Vol. 1, p. 1.

[49] S. Hussain and S. Mahmood, in "Advances in Plasma Physics," edited by M. Baranger and E. Vogt (Plenum, New York, 1968), Vol. 1, p. 1.

[50] R. G. Gill, in "Advances in Nuclear Fusion," edited by M. Baranger and E. Vogt (Plenum, New York, 1968), Vol. 1, p. 1.

[51] R. Yoshino, S. Tokuda, and Y. Kawano, in "Advances in Nuclear Fusion," edited by M. Baranger and E. Vogt (Plenum, New York, 1968), Vol. 1, p. 1.

[52] S. I. Popel, S. V. Vladimirov, and P. K. Shukla, in "Advances in Plasma Physics," edited by M. Baranger and E. Vogt (Plenum, New York, 1968), Vol. 1, p. 1.

[53] K. Roy, A. P. Misra, and P. Chatterjee, in "Advances in Plasma Physics," edited by M. Baranger and E. Vogt (Plenum, New York, 1968), Vol. 1, p. 1.

[54] ChRozina et al, in "Advances in Plasma Physics," edited by M. Baranger and E. Vogt (Plenum, New York, 1968), Vol. 1, p. 1.

[55] P. A. Andreev, F. A. Asenjo, and S. M. Mahajan, "On a consistent macroscopic description for a spin quantum plasma with interparticle interactions," arXiv preprint arXiv:1304.5780, 2013.

[56] B. Sahu, A. Sinha, and R. Roychoudhury, "Ion-acoustic waves in dense magneto-rotating quantum plasma," Physics of Plasmas, vol. 26, no. 7, 2019.

[57] R. W. Lorenz, "Linear and nonlinear coupling of waves in plasmas," Radio Science, vol. 7, no. 8-9, pp. 845–852, 1972.

[58] G. Manfredi, Fields Inst. Commun., vol. 46, p. 263, 2005.

[59] F. Haas, L. G. Garcia, J. Goedert, and G. Manfredi, Phys. Plasmas, vol. 10, p. 3858, 2003.

[60] Kumar P, Ahmad N, Laser and particle beams, vol. 38, pp. 159-164, 2020.

[61] Moslem, W.M., Ali, S., Shukla, P.K., Tang, X.Y., Rowlands, G., Phys. Plasmas, vol. 14, p. 082308, 2007.

[62] G. Chabrier, F. Douchin, and A. Y. Potekhin, J. Phys.: Condens. Matter, vol. 14, p. 9133, 2002.

[63] R. Ruffini, G. Vereshchagin, and S. S. Xue, in "Advances in Plasma Physics," edited by M. Baranger and E. Vogt (Plenum, New York, 1968), Vol. 1, p. 1.

[64] D. A. Uzdensky and S. Rightley, in "Advances in Plasma Physics," edited by M. Baranger and E. Vogt (Plenum, New York, 1968), Vol. 1, p. 1.

[65] P. R. Dip, M. A. Hossen, M. Salahuddin, and A. A. Mamun, The European Physical Journal D, vol. 71, pp. 1-8, 2017.

[66] A. Atteya, M. A. El-Borie, G. D. Roston, and A. S. El-Helbawy, Journal of Taibah University for Science, vol. 14, no. 1, pp. 1182-1192, 2020.

[67] W. Masood, A. M. Mirza, and M. Hanif, Physics of Plasmas, vol. 15, no. 7, 2008.

[68] W. F. El-Taibany, P. K. Karmakar, A. A. Beshara, M. A. El-Borie, S. A. Gwaily, and A. Atteya, Scientific Reports, vol. 12, no. 1, p. 19078, 2022.

[69] P. R. Dip, M. A. Hossen, M. Salahuddin, and A. A. Mamun, Journal of the Korean Physical Society, vol. 68, pp. 520-527, 2016.

[70] Pavel A. Andreev; Hydrodynamic and kinetic models for spin-1/2 electron-positron quantum plasmas: Annihilation interaction, helicity conservation, and wave dispersion in magnetized plasmas. Phys. Plasmas 1 June 2015; 22 (6): 062113.

[71] Mikhail V. Medvedev; Plasma modes in QED super-strong magnetic fields of magnetars and laser plasmas. Phys. Plasmas 1 September 2023; 30 (9): 092112.

[72] Z. A.Moldabekov, M. Bonitz, , & T. S. Ramazanov; Theoretical foundations of quantum hydrodynamics for plasmas. Physics of Plasmas, 25(3) (2018).

**Figure Captions**

Fig. 1: Variation of $k$ with $\omega$ for ions together with H=0 for $\theta = 4°$, $\Omega_0 = 0.0003$, $B_0 = 2 \times 10^{10} G$

Fig. 2: Variation of $\omega/\omega_{pe}$ with $kc/\omega_{pe}$ for electrons with H=0.75 (Non-dashed curve) for $\theta = 4°$, $\Omega_0 = 0.0003$, $B_0 = 2 \times 10^{10} G$ and H=0 (Dashed curve) with thermal velocity $V_{Te} = 10^7 \, cm/\sec$

Fig. 3: Variation of $\omega/\omega_{pp}$ with $kc/\omega_{pp}$ for positrons with H=0.75 (Non-dashed curve), $\theta = 4°$, $\Omega_0 = 0.0003$, $B_0 = 2 \times 10^{10} G$ and H=0 (Dashed curve) with thermal velocity $V_{Tp} = 1.05 \times 10^7 \, cm/s$

Fig. 4: Variation of $\omega/\omega_p$ with $kc/\omega_p$ for coupled e-p-i modes with H=1.1, $\theta = 4°$, $\Omega_0 = 0.0003$, $B_0 = 2 \times 10^{10} G$

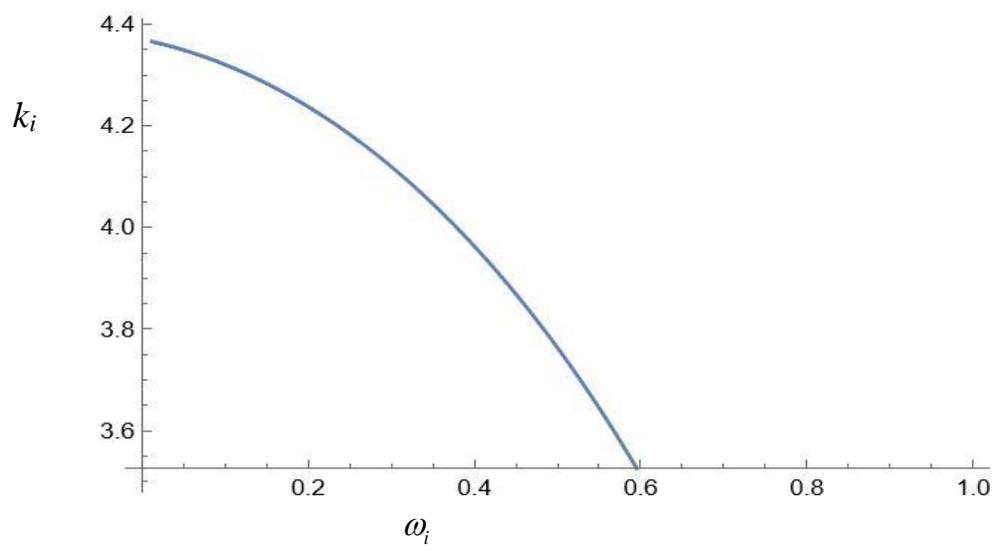

**Fig. 1**

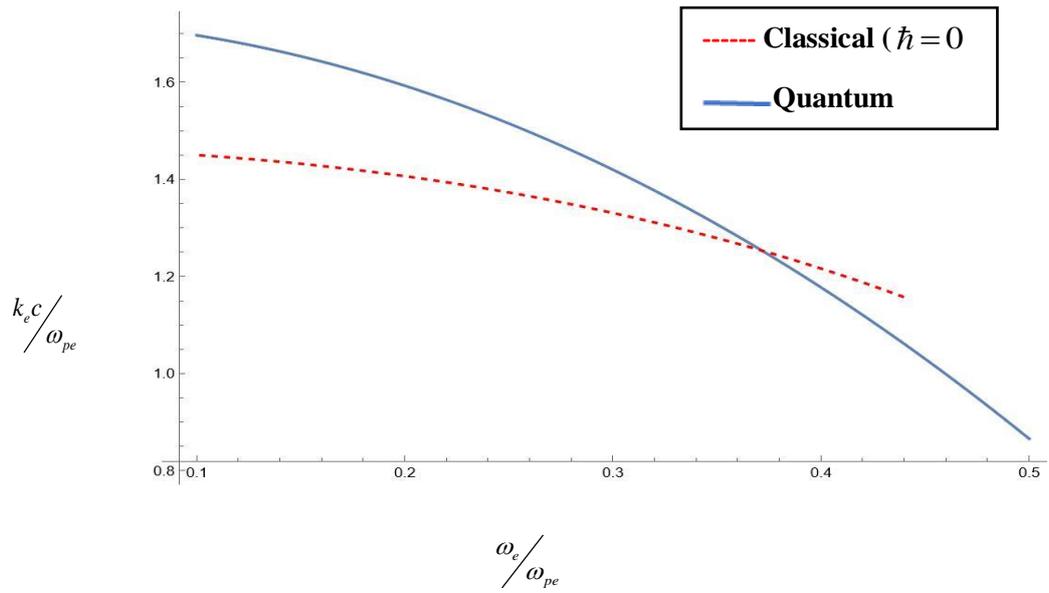

**Fig. 2**

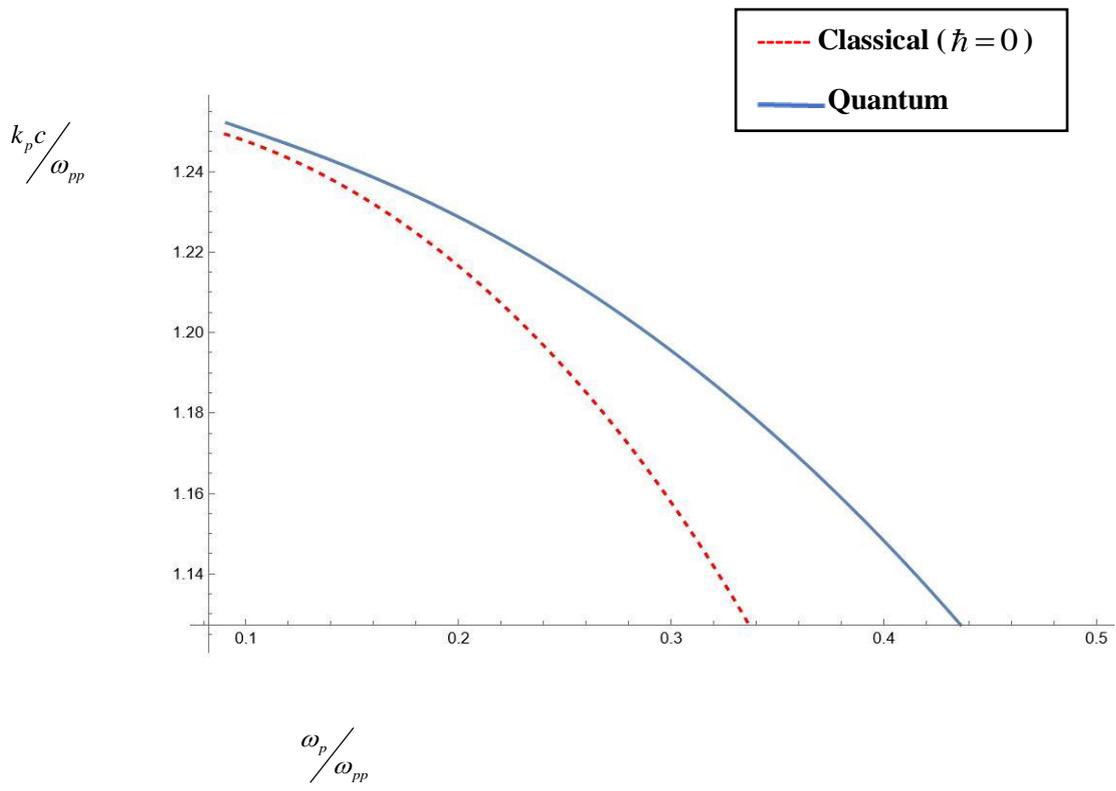

**Fig. 3**

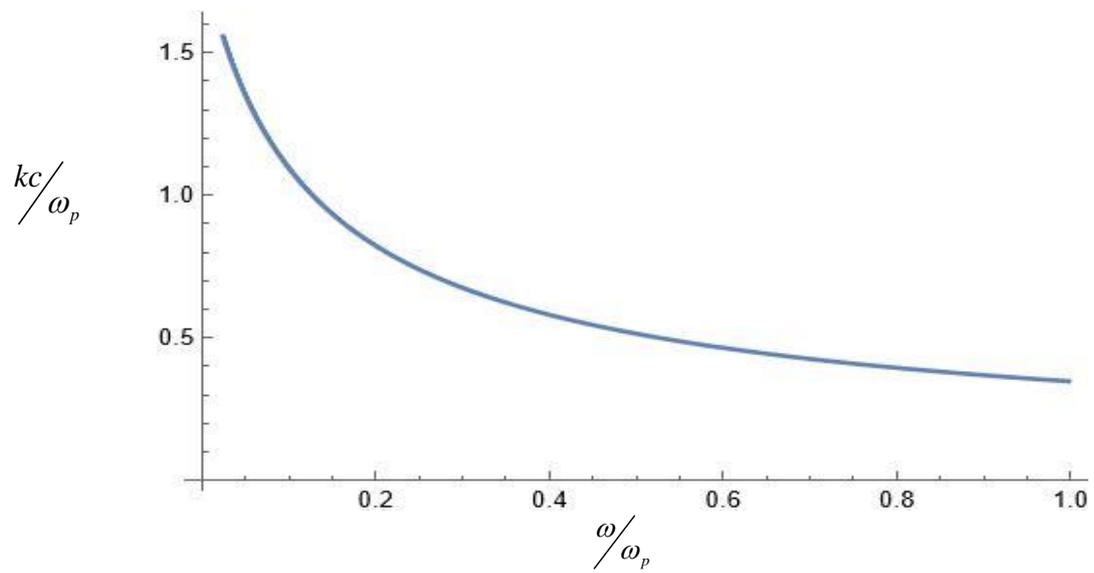

**Fig. 4**